\documentclass{article}
\usepackage{amsmath}
\usepackage{mathtools}
\usepackage{amssymb}
\usepackage{amsfonts}
\usepackage[numbers]{natbib}
\usepackage{relsize}
\usepackage{csquotes}
\usepackage{placeins}
\usepackage[margin=1in]{geometry}
\usepackage{amssymb}
\usepackage{csquotes}
\usepackage{hyperref}
\usepackage{graphicx,subcaption}
\usepackage{float}
\usepackage{color}

\renewcommand{\matrix}[1]{\mathbf{#1}}

\begin{document}

\title{Manifold Cities: Social variables of urban areas in the UK}

\author{
Edmund Barter$^{1*}$,Thilo Gross$^{1}$}

\maketitle

$^{1}$Merchant Venturers Building, Woodland Road, Bristol, BS8 1UB\\
$^*$ edmund.barter@bristol.ac.uk


\maketitle

\begin{abstract}
In the 21st century ongoing rapid urbanization highlights the need to gain deeper insights into the social structure of cities. While work on this challenge can profit from abundant data sources, the complexity of this data itself proves to be a challenge. In this paper we use diffusion maps, a manifold learning method, to discover hidden manifolds in the UK 2011 census dataset. The census key statistics and quick statistics report 1450 different statistical features for each census output area. Here we focus primarily on the city of Bristol and the surrounding countryside, comprising 3490 of these output areas. Our analysis finds the main variables that span the census responses, highlighting that university student density and poverty are the most important explanatory variables of variation in census responses.    
\end{abstract}


\section{Introduction}
In 2014 54\% of world's population inhabited cities~\cite{UNreport}. The United Nations project this fraction to rise to 80\% by 2050, amounting to an urban population of more than 6.5 billion~\cite{UNreport}. To visualize the increase in urban area, we can say that it is equivalent to building a city for 1 million people every week~\cite{AravenaTED}.

City life offers many opportunities, such as better chances of employment, a much wider range of cultural activities and greater diversity of people. However, it also has some disadvantages, such as greater problems with crime, congestion, lower air quality and higher living expenses. 
Many studies suggest a healthy social fabric plays a key role in emphasizing the positives of urban living~\cite{jacobs1961death}. Studies also show that social cohesion can be affected by design decisions, such as street layout and traffic restrictions~\cite{Appleyard1976,Appleyard1980}.

To analyze a city one can distinguish between the hard properties that are precisely defined and centrally regulated (e.g.~physical infrastructure, traffic planning, and zoning) and soft properties that arise from social interactions of the inhabitants (e.g.~social cohesion, the spirit and vibe of a city). The soft properties play an important role in general well-being \cite{kim2014perceived}, disaster response \cite{patterson2010role}, and the economy's ability to attract and retain qualified staff \cite{begg1999cities,fujita2001spatial}. Thus optimizing the hard properties that can be directly regulated to achieve an optimal development of the soft properties is a valuable target. 

Mathematical and computational models can help to understand how soft properties arise from the substrate of hard properties. They can thus help to elucidate the functioning of a city and aid decision makers \cite{Batty1971}.
Historically, models that include a very large number of variables have been used predict the effects of perturbations to the system, for example changes to transport infrastructure\cite{Batty1971,acheampong2015land}. These predictions can be very accurate, but rely on fitting for a large number of parameters.  
By contrast, models of \emph{urban economics} use simple rules to model a small number of variables and reproduce particular qualitative characteristics common to many cities. For example the emergence of multiple centers of economic activity separated by residential areas\cite{Fujita1982,Louf2013,Krugman,Dymski1996}. In these models, spatial heterogeneity emerges from the individual decisions of the occupants. However, the full diversity observed in real cities is not replicated and the predictions that extend from these models are usually not applicable for informing specific interventions or policies\cite{Barthelemy2016}.

It has recently been suggested that a new type of model of cities is both possible and desirable\cite{Barthelemy2016}. By taking a physics-inspired approach it may be possible to develop dynamic models that consider a small set of carefully selected aspects and then formulate qualitative or quantitative predictions based on the analysis of a small number of variables. This approach is particularly tempting as results do not emerge from a black box, but are gained through understanding the dynamics of processes. While no such model will be able to represent a city with its many subsystems in its entirety, a combination of different models could paint a comprehensive picture. The hope is thus that lightweight models that can be analyzed and understood in detail will eventually form the building blocks of a much coveted "science of cities"\cite{Batty2011}. 

A central challenge on the route to modelling the emergence of soft city properties is to identify a set of variables that capture these properties. The modeller is faced with a myriad of possibilities in which cultural and economic aspects of the city can be quantified. Clearly many of the variables that can be measured, such as household income, household size, employment status etc. have strong interdependencies. The challenge is therefore to pick an, ideally small, number of variables that represent the state of an area without creating unnecessary redundancy.

Selecting a specific set of variables can introduce bias to a model. If we pick specific variables that we believe to play some role in a given phenomenon, we ultimately come up with a model explaining the phenomenon in terms of these variables, potentially missing deeper causal connections. It is therefore desirable to pick the variables of a model in an objective way. Ideally we select variables according to their explanatory power, such that each variable that is added to the model maximally increases our power to describe differences in the city. This yields a model that can describe as much of the variation in the city with the given number of variables. Moreover, it is reasonable to assume that by focusing on the most explanatory variables we also identify the most important causal factors.    

An example is the analysis of the UK census general statistics, which provide a spatially resolved map of 1450 variables across the UK. To cluster the area into larger geographical groupings the UK's Office of National Statistics found it necessary to reduce the variables that are considered down to 60 \cite{ONS2015}. The selected set reflects an expert's opinion on the most important determinants of socio-economic structure.
However, this approach is undesirable as it disregards most of the available data, constitutes an additional source of bias and consequently reduces the quality of the results. There is thus a strong need for a principled mathematical approach to the construction of the socio-economic variables, that avoids introducing additional bias and takes all the available data into account.  

In this paper we address the challenge of constructing the leading explanatory variables from spatially resolved social data. Starting with the UK 2011 census dataset we consider the city of Bristol and its surroundings including several smaller settlements. The data is analysed using the diffusion map method \cite{Coifman2005,Coifman2006}. This method uses the physical intuition and mathematical results on diffusion processes to identify explanatory variables. We emphasize that, in contrast to related approaches such as previous work on spectral clustering results \cite{cranshaw2012livehoods}, we do not consider a diffusion process in physical space, but data space. We demonstrate that this can be used to gain an understanding of the spatial organization of the city and also identifies a set of variables that can be used in future models of British cities. We analyse the first two of these variables in detail and show that they capture the distribution of the university student population and poverty as leading factors that shape the variation in the census response.    


\section{The 2011 Census data}
In the UK censuses of the complete population began in 1801, and have occurred every 10 years since (with the exception of 1941)~\cite{censusHistory}. While early censuses involved simple head counts, the most recent UK census, conducted on 27$^\mathrm{th}$ March 2011, collected a wide range of demographic and economic data about households and their residents, including their ethnicity, type of home ownership and employment details. 

The results of the census are published in a mildly aggregated form where responses from approximately 100 households are combined into so-called \emph{output areas} (OAs). OAs are constructed after the data has been collected. They are drawn such that there is a minimum number of 60 households in each OA, the OAs are spatially contiguous, and the homogeneity of responses in an OA is maximized. In space, the OAs form a complete tesselation of the UK's land surface, which means that areas without residential population, including public parks, industrial estates etc., are grouped with adjacent residential areas to form OAs. In effect these rules mean that OAs in cities can be very small comprising only a single building or part of a building, whereas OAs in the countryside can be up to three orders of magnitude larger. 

The census results for England and Wales is a dataset containing census responses for 175434 OAs. Here we consider particularly the data for Bristol and the surrounding countryside, that is the area of the Local Authorities of Bristol, North Somerset, Bath and North East Somerset, and South Gloucestershire. This combined area of Bristol and the surrounding Local Authorities comprises 3490 output areas, covering 1331$\mathrm{km}^2$.


\begin{figure*}[htb]
\centering
{\includegraphics[width=0.6\textwidth]{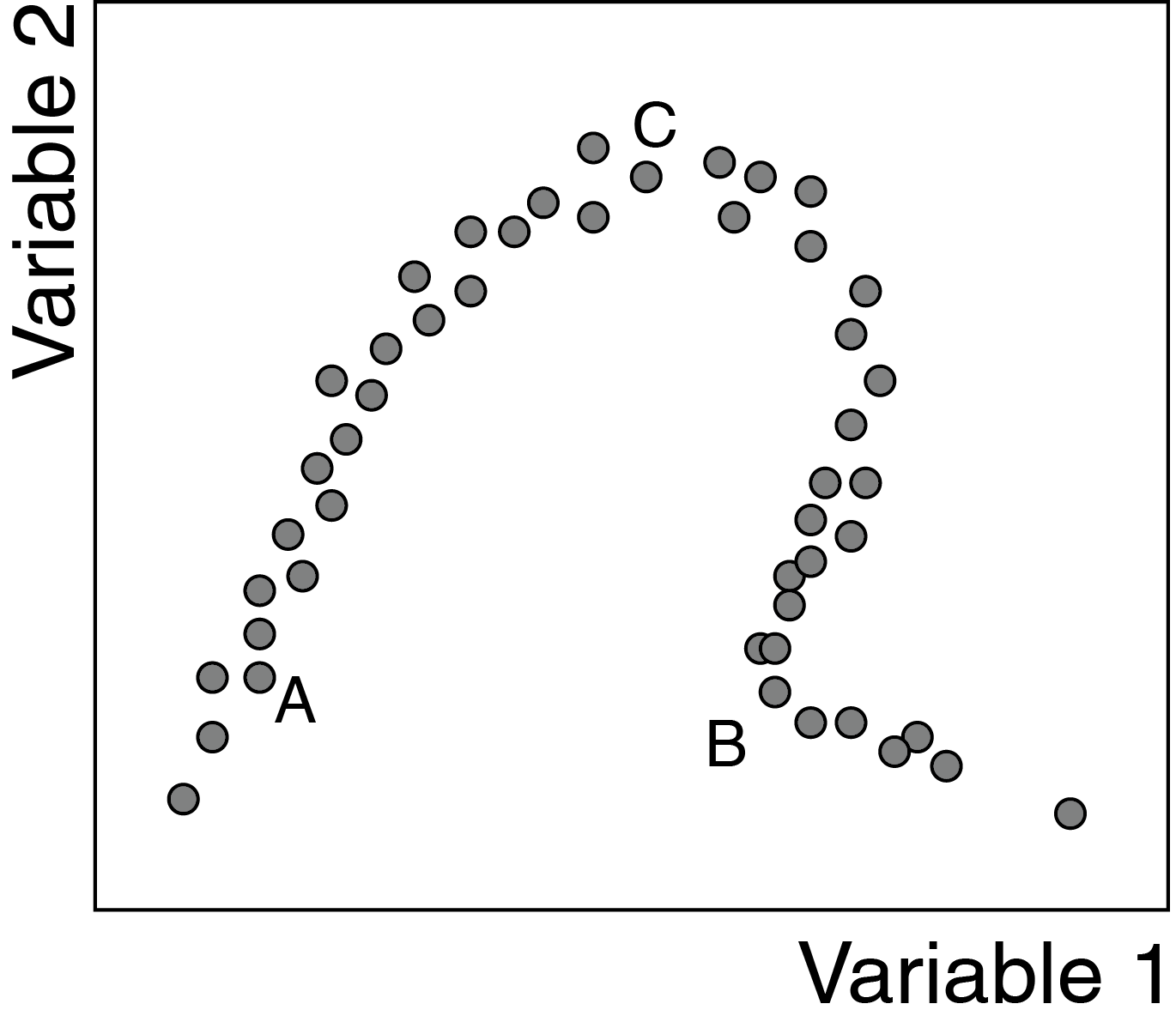}}
\caption{ Illustrative sketch in 2D. While the data space is two-dimensional the data points (circles) cluster around a one-dimensional manifold. While the Euclidian distance between points may be a good way to quantify close points it fails on larger scales. For example point A seems to be closer to B than to C in terms of Euclidian distance, but quantifying distance in terms of distance along the manifold (which is likely the actual distance A would have to travel to 'become like B') reveals that A is closer to C than to B.}
\end{figure*}

\section{Diffusion maps find hidden manifolds in data}
The variables of census data span a 1450-dimensional space. We can visualize every OA as a point in this space. If the data were two-dimensional we could simply plot the points in the plane to reveal how the points are distributed in the underlying space. For example we might find that points cluster around a one-dimensional curve (Fig.~1). In that case a more natural way of describing the position of an OA is to specify how far along the one-dimensional curve it is located. 
In high-dimensional data such underlying structures also typically exist \cite{Coifman2005,Coifman2006}. They may be points, curves, or volumes, which (mathematically speaking) we can summarize under the term manifolds.  

A simple method sometimes used to approximate manifolds in a dataset is principle component analysis (PCA) \cite{Pearson1901}: Given the original data space we identify the directions along which the greatest variation in the data occurs. Thus we find a set of variables that quantify the location of a point in the directions of highest variation. Although widely used, PCA has some notable drawbacks. Most prominently, it can only find linear structures in the data, e.g.~straight lines, flat planes, etc. Hence principle component analysis would not produce the desired results in the example from Fig.~1. 

A less widely known, but perhaps more important limitation of PCA and several related methods is that they rely on a global metric: The distances between all pairs of data points enter the calculation, even if a given pair of data points is very far apart. This is problematic because the euclidian metric, which is often assumed, is typically only appropriate locally. For illustration consider Fig.~1. According to euclidian distance A-B is shorter than the distance A-C. However, if the points in the figure represented different OAs, the totality of the data suggests that if A would develop in time to approach a state close to B it must pass close to C in the process. Thus in a more natural metric the distance A-B is longer then the distance A-C. Such uncertainties constitute a source of noise in the analysis. The effect of this noise gets more severe with increasing dimensionality of the data  \cite{ONS2015} obscuring the salient information; a phenomenon known as the curse of dimensionality.     

To address the issues above, Coifman et al.~\cite{Coifman2005,Coifman2006} proposed the diffusion map, which we revisit in the remainder of this section. The diffusion map is a manifold learning method that does not use information from distant pairs. We start by computing the distances between all pairs of nodes, but then discard distance measurements that are too large to be trusted, according to some properly chosen rule.   
The result of the procedure above can be visualized as a network where nodes are data points and weighed links indicate the proximity of data points that are mutually close. Several approaches are available for exploring the network using a geometric embedding~\cite{estrada2014hyperspherical,grover2016node2vec}. Diffusion maps discover its manifolds using a harmonic analysis. For this purpose one computes the eigenvalues and eigenvectors of a Laplacian matrix describing the network. This matrix also governs the dynamics of random walks on the network, hence the name \emph{diffusion map}. As a result of this procedure one obtains the Laplacian eigenvectors and eigenvalues. The eigenvectors corresponding to the smallest positive eigenvalues encode directions of large variation: Each vector has the dimensionality of a dataset, for example in our set of 3490 OAs, each eigenvector has 3490 entries, i.e.~one for each output area. Thus an eigenvector assigns a new variable to each OA. As shown in \cite{Coifman2005,Coifman2006} these variables indicate the location of an OA along the manifolds present in the data.  

In summary the analysis used in this paper consists of the following steps, which we now describe in more detail:
\begin{enumerate}
\item Standardize the data
\item Compute Euclidean distances between data points
\item Construct a similarity matrix
\item Threshold the similarity matrix
\item Define a Laplacian matrix
\item Use the spectral properties of the Laplacian to map data points into low-dimensional space spanning the manifolds
\end{enumerate}
We begin with the data in a $M\times N$ matrix $\bf A$, where $M=3490$ is the number of OAs and $N=1450$ is the number of census variables for each point.

First, we standardize all the columns of $\matrix{A}$, so that each variable has a mean of zero and a standard deviation of $1$. This is to ensure each variable is considered on the same scale in the dataset. We obtain the standardized data matrix $\matrix{\hat{A}}$, the components of which are
\begin{equation}
\hat{A}_{mn}=\frac{A_{mn}-\mu_n}{\sigma_n},
\end{equation}
where
\begin{equation}
\mu_n=\frac{\sum_m A_{mn}}{M},
\end{equation}
and
\begin{equation}
\sigma_n=\sqrt{\frac{\sum_m (A_{mn}-\mu_n)^2}{M}}
\end{equation}
are, respectively, the mean and standard deviation of the $n^{\mathrm{th}}$ column of $\matrix{A}$.

Second, each row of $\matrix{\hat{A}}$ contains the coordinates of an OA in a $N$-dimensional space. We then define a $M\times M$ distance matrix $\matrix{D}$, such
that 
\begin{equation}
D_{ij}=\sqrt{\sum_n (\hat{A}_{in}-\hat{A}_{jn})^2},
\end{equation}
is the euclidean distance between OAs $i$ and $j$ in the data space. 

Third, we convert the distances between data points to similarity scores. We define the similarity $C_{ij}$ between two data points $i,j$ as   
\begin{equation}
C_{ij}=k(D_{ij}),
\end{equation}
where $k$ is a properly chosen kernel, here $k(d)=1/d$. The diagonal elements $C_{ii}$ are set to zero. The resulting similarity scores $C_{ij}$ form a $M\times M$ matrix $\matrix{C}$, in which high values now indicate close similarity between the respective OAs.   

Forth, we threshold the data by setting most of the small entries of the similarity matrix $\matrix{C}$ to zero. As it is hard to define an absolute threshold of similarity we use a heuristic procedure in which an entry $C_{ij}$ is kept if it is among the top-10 highest similarity scores for either OA $i$ or OA $j$. Otherwise it is discarded and set to zero. This ensures that each node in the network has at least $10$ links, while removing a large number of weak, low-similarity links, avoiding the curse of dimensionality, described above. An additional benefit is that we are left with a sparse network that is numerically more efficient to process. 

Fifth, we construct the normalized Laplacian matrix, an $M \times M$ matrix defined by 
\begin{equation}
L_{ij} = \left\{ \begin{array}{l l} 1 & \hspace{2em}\text{for $i=j$} \\  -c_{ij}/\sum_n c_{nj} & 
\hspace{2em}
\text{otherwise.} 
\end{array}\right. 
\end{equation}
This matrix governs dynamics of a discrete-time diffusion process of the network and is closely related to the Laplacian matrix that describes, for instance, the response of a network of masses and rods to a mechanical perturbation.  

Sixth, we explore the structure of the dataset by computing the eigenvalues and eigenvectors of $\bf L$. Laplacian matrices are positive semi-definite matrices and thus the eigenvalues are either positive or zero. The number of zero eigenvalues is identical to the number of components in the network of data points. Thus there is always at least one zero eigenvalue. The corresponding eigenvector does not carry information and can be disregarded. If more than one zero eigenvalue exists the network has become disconnected in the thresholding step. In this case it is best to repeat the analysis with less aggressive thresholding. In the investigation presented here this was not the case. 

Of particular interest are the positive eigenvalues that are closest to zero as their corresponding eigenvectors span the main directions of the manifolds in the data. In the following we focus particularly on the two eigenvectors with the smallest non-zero eigenvalues.   

\section{Results}
We start by considering the smallest positive, and hence most important, Laplacian eigenvector. As this vector assigns an entry to each of the OAs we can visualize the eigenvector as a spatial map where the results are colour-coded from the most positive to the most negative entry (Fig.~2A). 

\begin{figure*}[htb]
\centering
{\includegraphics[width=0.395\textwidth]{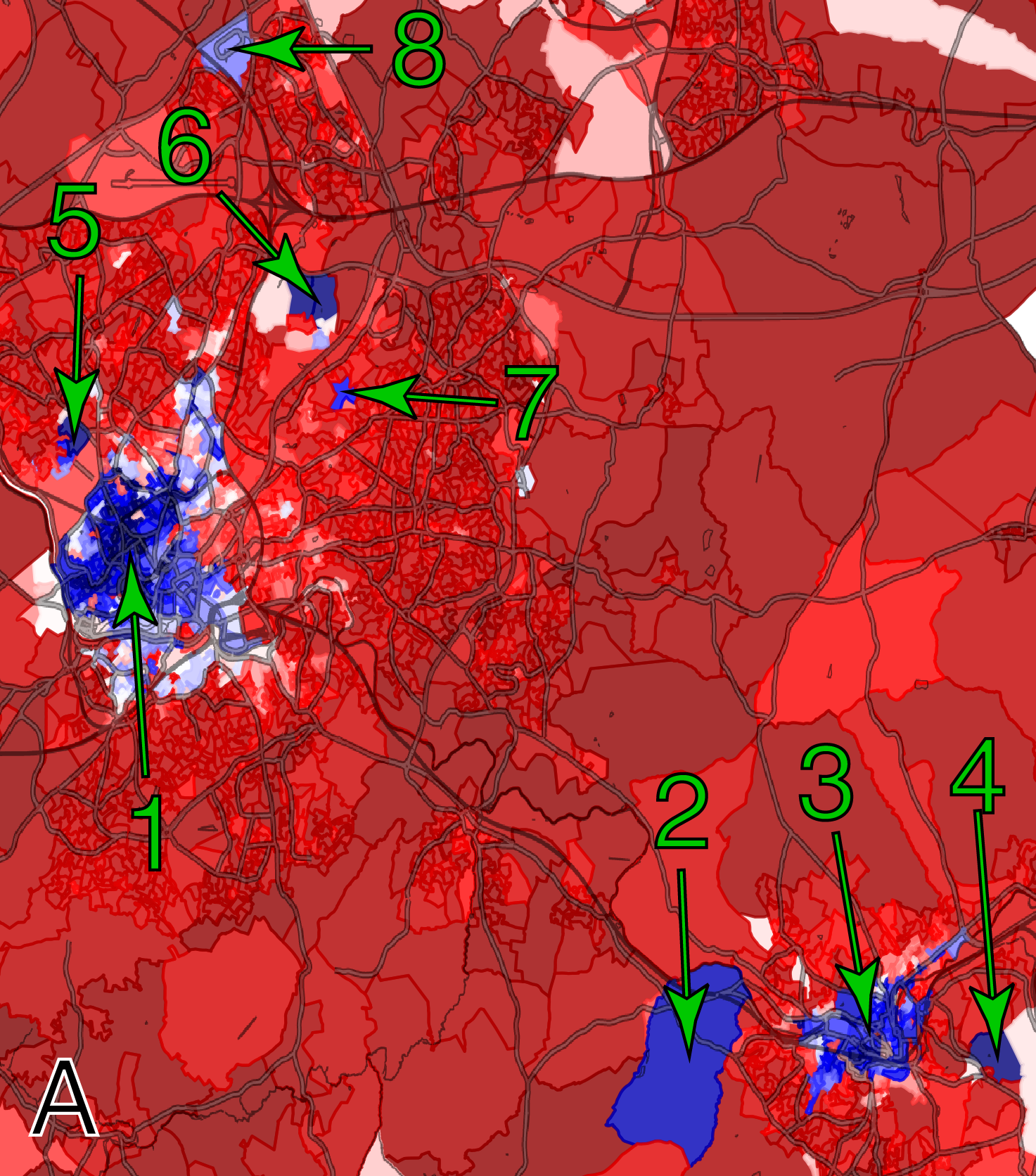}}~
{\includegraphics[width=0.595\textwidth]{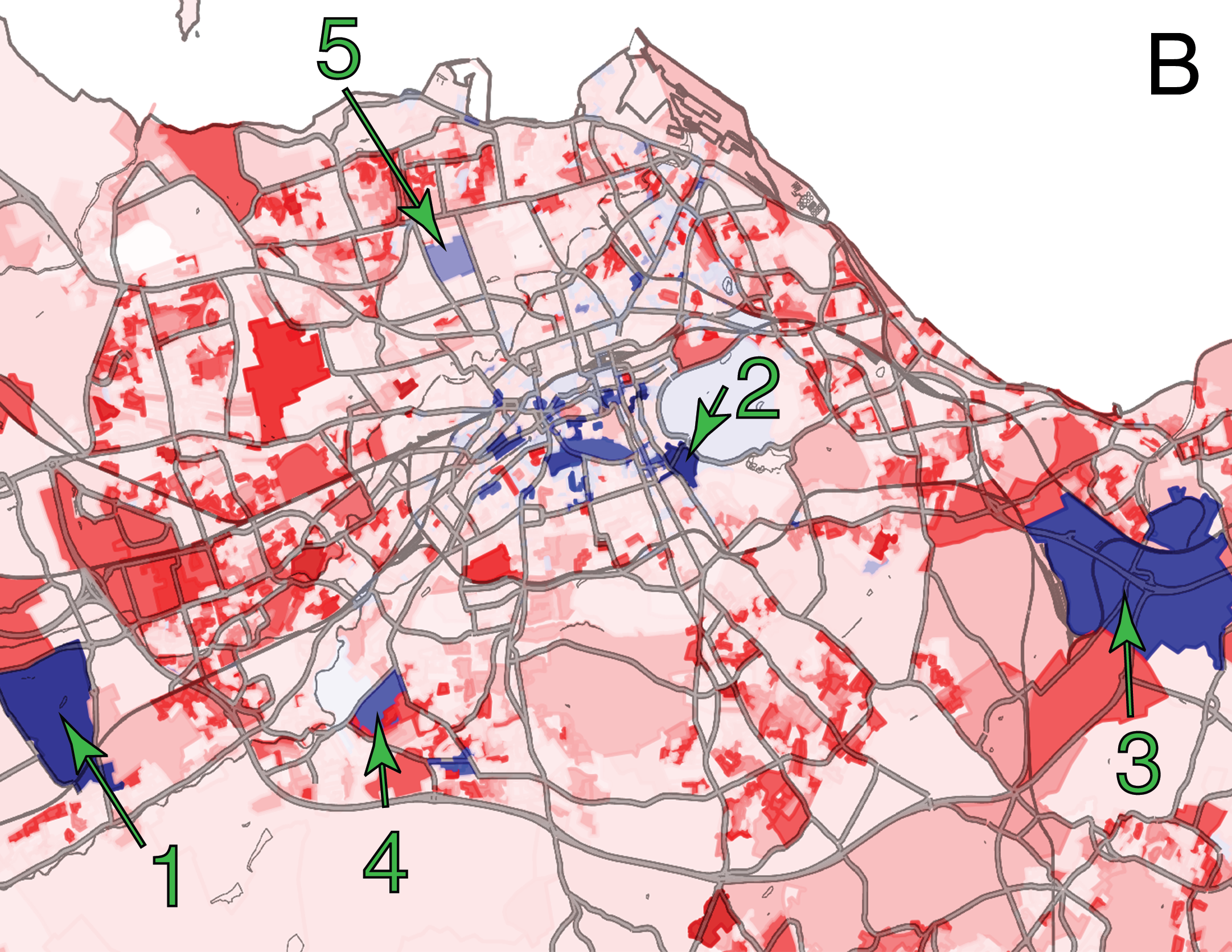}}
\caption{ Diffusion eigenvectors locate the student population in Bristol and Edinburgh. A: The smallest positive eigenvector of the diffusion map is plotted on a map of Bristol using a color scale from red (most negative entries) via white to blue (most positive entries). High values of the eigenvector occur in areas with a dense student population including University of Bristol and Bristol city centre (1), Bath Spa University (2), Bath city centre (3), University of Bath (4), University of Bristol Halls of Residence (5), University of West of England Frenchay Campus (6), University of West of England Glenside Campus (7), Aztec West technology park (8). B: Map of Edinburgh coloured in the corresponding coulour of the most similar output area in Bristol. Again blue areas coincide with dense student population, including Heriott-Watt University (1), University of Edinburgh (2, points at central halls of residence but university buildings are spread through a wider area nearby consistent with the eigenvector), Queen Margaret University (3), Ministry of Defence University Officers Corps site (4), and to a lesser extent Fettes College (5). Lighter colours in Edinburgh suggest the student population is less concentrated than it is in Bristol.}
\end{figure*}

We first note that this eigenvector is \emph{localized} in certain OAs: There are few areas (blues) where the eigenvector has a strong positive entries, while elsewhere (reds) it is close to zero. Moreover, the OAs where the eigenvector has a strong positive entries occur mainly in 6 large but relatively sharply delineated patches, with the largest patch containing $>$100 OAs. 

The existence of large contiguous patches is the first mild indication that the diffusion map picks up salient features of the dataset. Because no information on the spatial location of patches or neighborhood relationships between them is used, random or strongly noisy results would lead to a random sprinkling of red and blue. 

One limitation of the method is that it is not intuitive how the eigenvector emerges from the original census variables. This limits our ability to attribute a meaning to the signal picked up by the eigenvector. However, a human-intuitive meaning is discoverable if we take additional information into account. 

The six main patches that we identified show a very distinct pattern. Three of the six patches coincide with university campuses, which are the University of West of England's main campus at Frenchay, the University of Bath and Bath Spa University. One additional patch coresponds to Aztec West, a large technology park that maintains close links to regional universities. The remaining two patches are the city centre area of Bristol, which is also home to the University of Bristol, and the city centre of Bath where some departments of the University of Bath are located and which has a high residential student population. 

We conclude that the first eigenvector found by the diffusion map is localized on areas that have a large residential population of university students. We corroborated this finding by examining in detail some OAs which correspond to moderate eigenvector entries (light blue) while not being directly part of one of the bigger clusters. All of these were also found to contain University related infrastructure, such as outlying Halls of Residence of the University of Bristol and the University of West of England's smaller Glenside campus. Moreover, all of the Univerity's Halls of Residence fall within blue areas. We also verified that the centre of the nearby city of Weston-super-Mare, which does not have a local University does not receive any significant entry in the first eigenvector (not shown). 

So far we have presented evidence for the hypothesis that the most important variable extracted by the diffusion map is the density of university students. A notable
drawback was that that this was not possible using automatic methods to arrive at this human-intuitive interpretation. Visual inspection of maps and local knowledge was required to find that the eigenvector identified universities. However, having local knowledge in one city (Bristol) we can now use the information about the manifold in the data to try to locate the student population in a different city. 

As an example we consider the city of Edinburgh, Scotland. The Scottish census is largely similar to the census in England an Wales, but some questions are presented differently and subtle differences also exist in the data management and aggregation. We confirmed that if we rerun the diffusion map analysis we obtain a result that is very similar to the results as for Bristol. However, here we want to make the point that the results from Bristol can actually be used to understand the structure of Edinburgh. For this purpose we selected 231 variables by the criteria that their identifying names in the the England and Wales, and Scotland census could be easily matched. We then assign to each OA in Edinburgh a value that is identical to the eigenvector entry of the OA in Bristol to which it is most similar. Similarity is quantified by Euclidian distance in terms of the 231 selected variables, such that not every Bristol OA must have a match in Edinburgh and multiple Edinburgh OAs can be assigned the value from a single Bristol OA. 

Plotting the result on the map of Edinburgh (Fig.~2B) again reveals some localized areas of very high entries (blue). Based on the results from Bristol (not using local information from Edinburgh, apart from the census) our hypothesis is that these must be the areas with the densest student population. We can now use local information to confirm this hypothesis. The OAs in Edimburgh that receive the highest entries are on the campuses of Heriot-Watt University, the University of Edinburgh, and Queen Margaret University. Other areas that receive significant entries are on lively streets close to the University of Edinburgh, featuring a mix of residential and university buildings; a ministry of defence site on which a university officers training corps is located and an associated barracks area. 

We also see some signal from the university eigenvector at Fettes College, a large private boarding school. This serves as a reminder that the diffusion map does not pick out any characteristic that is directly reflected in the census but a complex nonlinear combination of different statistics, including age bracket, occupational status, household size, and many more. We were not able to confirm if large populations of boarding school students get consistently picked up by the eigenvector as other similar schools in the area are located in heavily university dominated OAs. 

\begin{figure*}[htbp]
\centering
{\includegraphics[width=0.45\textwidth]{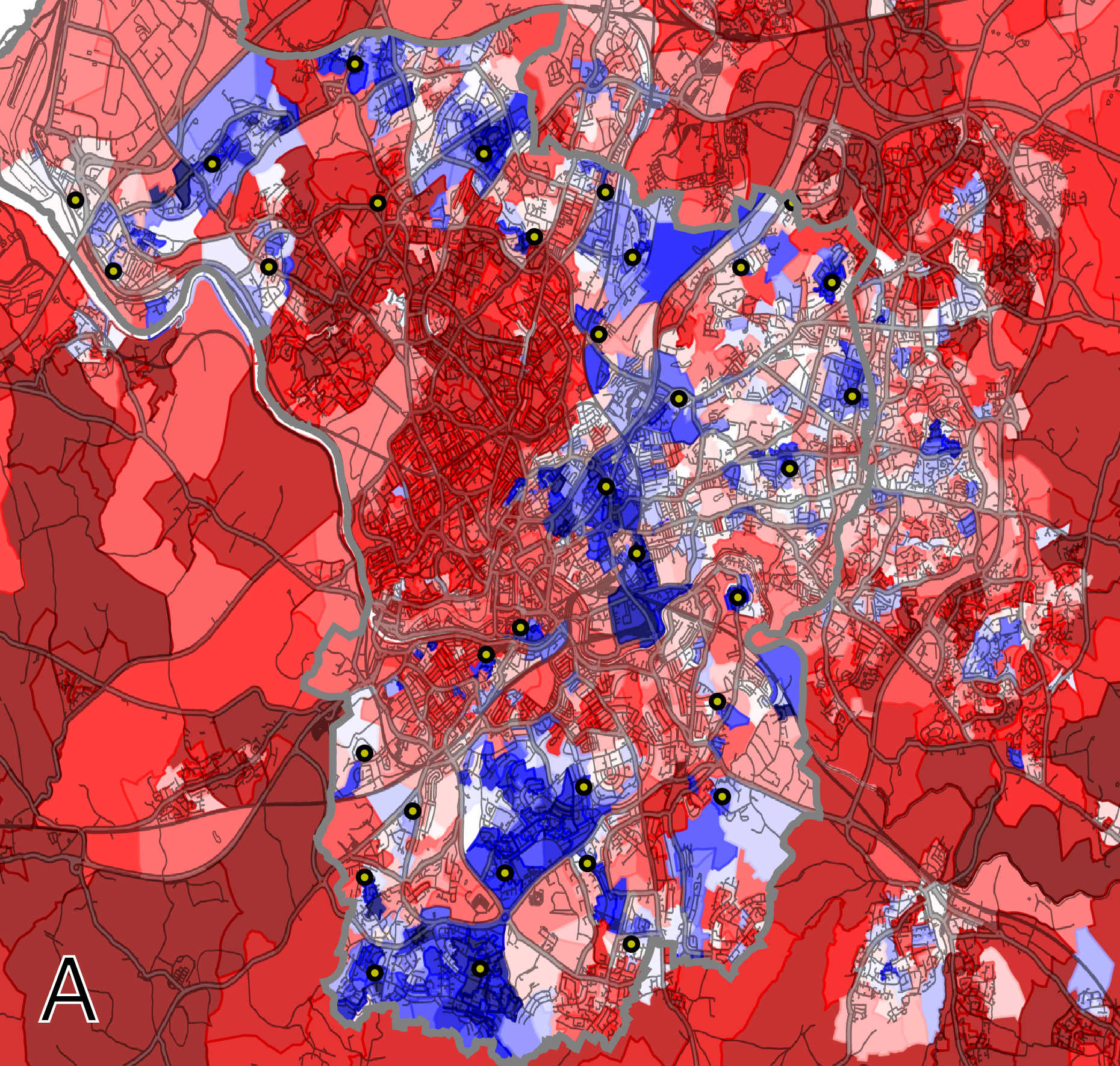}}~
{\includegraphics[width=0.45\textwidth]{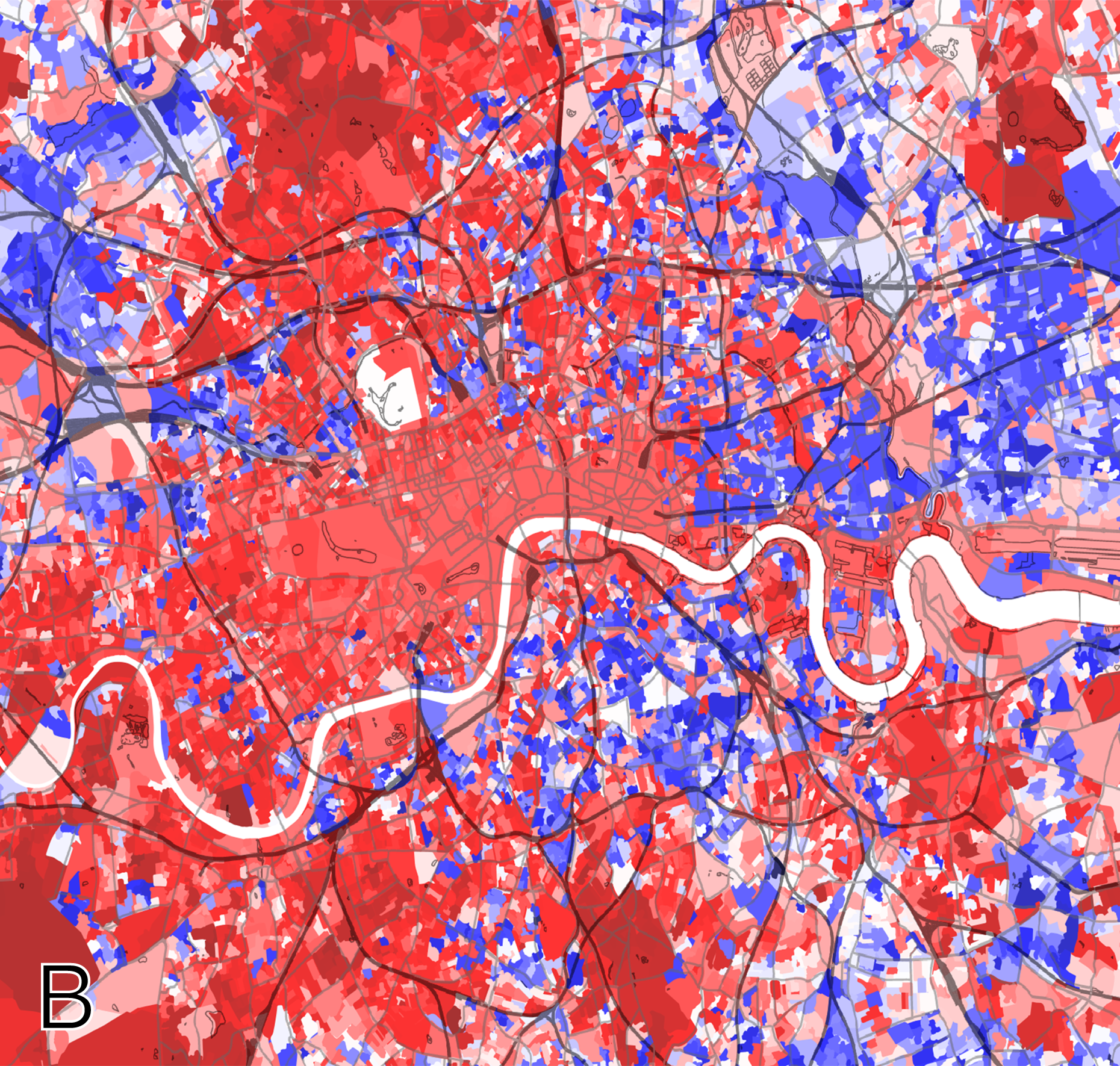}}\\
{\includegraphics[width=0.45\textwidth]{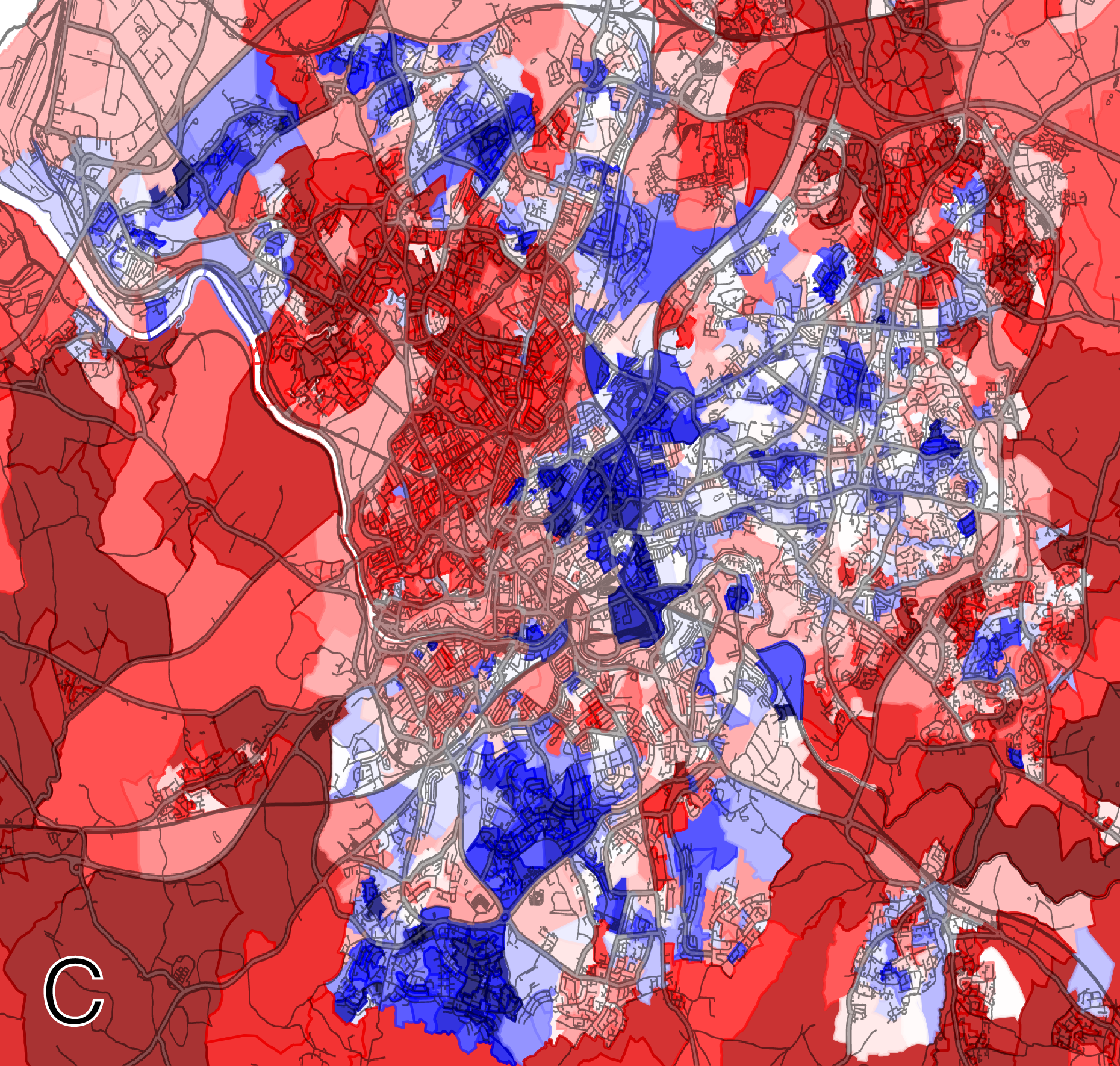}}~
{\includegraphics[width=0.45\textwidth]{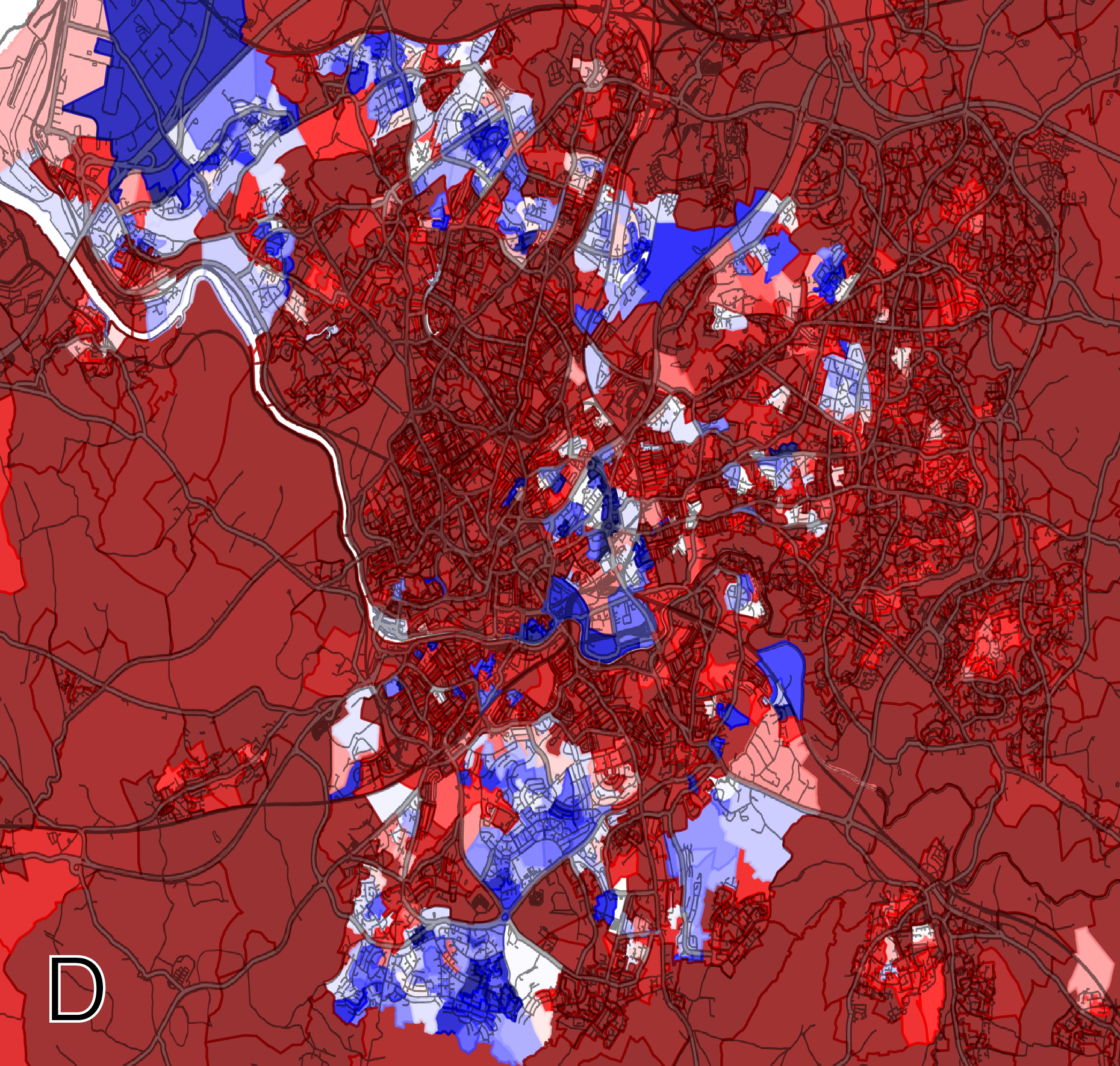}}\\
\caption{The second diffusion eigenvector is an indicator of poverty. A: Color-coding the second smallest positive diffusion eigenvector in Bristol highlights areas (blue) that coincide locations of social housing estates (yellow dots). Dots are based on a report that covered only the Bristol unitary authority area (grey border), hence social housing estates in the rightmost part of the map are picked up by the eigenvector but not marked with a dot. B: OAs in London colored with the color of the most similar Bristol area reveal an intricate pattern of rich and poor. C: Rerunning the Bristol analysis without information on the number of households renting from council reveals an eigenvector that is virtually unchanged. D: By contrast plotting only the proportion of people living in social housing reveals a rougher picture, that shows a similar overall pattern with some notable differences. Detailed analysis of the discrepancies in areas such Knowle West (center of bottom third of the map) shows that the eigenvector is an indicator of economic deprivation that largely but not always coincides with council housing. 
}
\end{figure*}

We now turn to the eigenvector corresponding to the second smallest positive eigenvalue. Plotting this eigenvector reveals localization on a complex pattern of areas (Fig.~3A). 

Based on local knowledge we noted that the distribution of high-eigenvector-entry areas correlate with the location of major council estates. In the UK council estates are, often purpose built, social housing projects that provide affordable accommodation citizens that are not able to afford housing at the market rate. In Bristol council estates take a wide variety of forms ranging from clusters of 1970s high rise buildings to small semi-detached houses, originally built for US airmen in the 1940s.   

To confirm the hypothesis that the second eigenvector (and hence second most important explanatory variable) in the Bristol census data is broadly related to poverty, we compared the pattern from the diffusion map with a list of major estates from a council report \cite{Malpass2005}. While this report covers only the Bristol unitary authority area and thus a smaller area than the eigenvector the correspondence between identified estates and the areas identified by the diffusion map is very good.

\begin{figure*}[htbp]
\centering
{\includegraphics[width=0.9\textwidth]{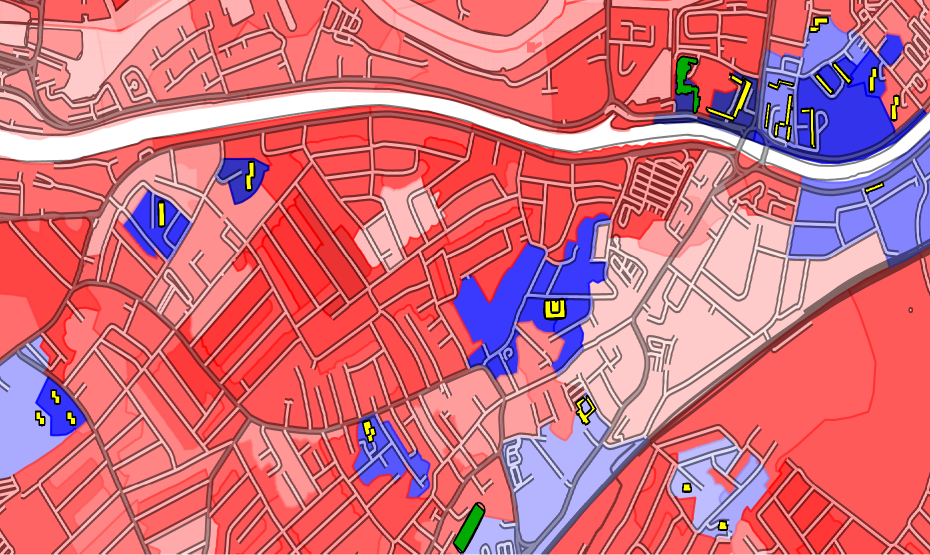}}
\caption{Zooming  into  smaller  areas  reveals that for larger residential buildings the eigenvector offers almost building level accuracy. Areas highlighted by the eigenvector (blue) coincide well with large social housing complexes (yellow outlines).  Also  a  hospital  and  a  social  home  for  the  elderly  (green  outlines,  top  and  bottom,respectively) are highlighted by the eigenvector
}
\end{figure*}

Because council housing is widely recognized to be a major factor in the social fabric of British cities this information is captured directly by the census, hence the number of  household `renting from council' (RFC) is one of the 1450 columns in the dataset. On the one hand this allows us to explore in greater detail if the second eigenvector is an indicator of poverty. On the other hand it raises the question if the diffusion map in this case only picks out information that is straightforwardly provided in one column of the data matrix. 
To answer these questions we reran the diffusion map analysis on the Bristol data after removing the RFC information from the dataset (Fig.~3C). In a separate plot we show also  the proportion of RFC households directly from the census data (Fig.~3D). 
Comparing the original second eigenvector (Fig.~3A) and the modified eigenvector without RFC (Fig.~3C) reveals that the eigenvectors remain almost unchanged. Thus the result from the diffusion map is not strongly dependent on the RFC column being present. Nevertheless
comparing the eigenvectors with the proportion RFC (Fig.~3D) shows a strong resemblance between the eigenvectors and households RFC. Some notable differences exist, of which we would like to discuss one in some more detail as we judge it to be the most instructive. 

In the Knowle West area of Bristol the data show that only a moderate proportion of households are RFC. However, Knowle West is one of the areas that is most strongly highlighted by the diffusion map.  The area was a modestly prosperous working class neighbourhood until the closure of a local factory in 1990 led to the loss 5,000 jobs immediately, and an estimated further 20,000 through cascading effects \cite{knowleWest}. In the early 2000s the area was widely recognized as an example of economic deprivation. It received extensive coverage in national news and provides the backdrop for the novel Shawnie \cite{trewavas2006shawnie}, highlighting the link between deprivation and ant-social behaviour.         
At the time of the census some properties in the area had been sold to the council to provide social housing, while many other families remained in privately-owned accommodation, but suffered from widespread unemployment. Only considering the housing status thus misses the intense economic deprivation existing in this area that is detected by the diffusion eigenvector. 

We conclude that the second diffusion eigenvector is an indicator of poverty, in the sense that the eigenvector detects a direction that spans a major dimension of the census manifold in data space for which poverty provides the right intuition. We emphasize that this dimension, identified by the algorithm, is a complex nonlinear combination of many different statistics reflecting much more of the census than just housing status or income level.   

To test the accuracy of predictions we explored a small area comprising the Southville and Bedminster neighborhoods of Bristol in greater detail. This area was chosen as it is very diverse and has experienced a rapid improvement in the years following the 2011 census. The area is also relatively close to the university which allowed us to explore it on foot and talk with residents about the history of individual buildings. We were thereby able to identify major council operated residential buildings (Fig.~4). Comparing the locations of these buildings to the diffusion eigenvector shows very good agreement. We verified that most of the highlighted areas outside the residential buildings is explained by green space and non-residential buildings which are grouped into the same OAs as council housing. Thus for large residential buildings the eigenvector identifies poverty almost to a building-level accuracy. 

Finally, we also used the results from Bristol to predict the pattern of poverty in central London. Similarly to the Edinburgh example above we coloured the OAs in the areas using the eigenvector entry using the most similar OA in the Bristol data. Because English census data was available in this case we defined similarity using the Euclidian distance in the full 1450 dimensional data space. The observed pattern meets the general expectations about the distribution of wealth in the British capital, such as the affluent neighborhoods in Mayfair and Belgravia, with more poorer areas south of the river Thames. Interesting smaller details include for example the strong differences between the rich financial district at Canary Wharf and adjacent economically deprived areas on the Isle of Dogs. Another contrast, also picked up by the diffusion map, is the difference between more deprived areas of the Golden Lane Estate and the Barbican, a rare example that is commonly described as a middle class council estate \cite{barbican}.  

\section{Discussions and Conclusions}
In this paper we used diffusion maps to find the social variables that explain the main variation in the census responses in the city of Bristol and the surrounding countryside. This demonstrates that diffusion maps can be used to discover major social dimensions of the urban fabric in freely available data.

Recently other notable methods have been presented that are alternatives to the harmonic analysis used to produce the low-dimension, geometric representations of network structure in diffusion maps~\cite{grover2016node2vec,pereda2019visualization}. Some of these advances are already being applied to analyze cities~\cite{akbarzadeh2018communicability}. We expect that these methods could be applied to census data to find results of similar quality. 

Our results show that density of university students and economic deprivation are the two most important explanatory variables spanning the manifold of census responses in a typical British city. These variables emerge as complex nonlinear combinations of many different statistics captured by the census. 

We also showed that the information on the manifolds learned from the Bristol data can be used to detect features such as university infrastructure and economically deprived areas in other cities, suggesting that the social variation observed in Bristol is sufficient to extract variables that are also informative for other British cities. 

The diffusion map is a particularly attractive method for these types of analysis because it offers high quality results and is based on a strong physical intuition that leaves almost no free parameters. In our analysis the only parameter that could be varied is the number of links per node that are retained in the thresholding. For this parameter clear constraints exist, because we want to reduce the number of links without disconnecting the network of datapoints. Within these constraints we found the predictions of the diffusion map to be reasonably robust. The diffusion map thus constitutes a method that allows analysis of census datasets without introducing any biases except those inherent in the selection of census questions . 

In the analysis of social data it is a persistent problem that human intuition is needed to interpret and verify results. A critic can thus argue that an intuitive result is not new, and an unintuitive one not true. While the method proposed here does not entirely solve this problem, some progress has been made:
Our results show that, independent of human interpretation, there are coherent structures in the census that act as major explanatory variables.
While human intuition is still necessary to attach the labels ``students'' and ``deprivation'' to these structures, the results show that the structures have a reality which is independent of these labels, and are not just a product of interpretation.     

Because we establish the reality of a structure in the census, attaching a specific label to this structure then amounts to formulating a testable hypothesis. 

Another advantage of the diffusion map is that it not only identifies but also ranks the discovered variables. Although one might have expected deprivation to feature prominently in this analysis (we did not), it is perhaps a surprise that we find that students to have an even greater explanatory power. We caution however, that this may be due to the census containing many question that would be answered similarly by students. In contrast to the students and deprivation, for example ethnic origin, did not stand out as among the major eigenvectors.   

In this paper we have focused almost entirely on the two leading eigenvectors identified by the diffusion map. Variables are identified by eigenvectors of the diffusion map because they induce coherent responses to census questions. Subsequently, we are convinced that many more eigenvectors provide salient information. However, the large amount of work required to produce and verify a hypothesis on the source of coherent responses for each eigenvector, means that their analysis exceeds the scope of the present paper. 

We conclude that there is a large untapped potential for diffusion map analysis of census data. Even within the Bristol dataset there many more eigenvectors remain to be explored. But the diffusion map methodology can easily scale to much bigger datasets, such as the entirety of the England and Wales census dataset. In addition, similar rich and finely resolved census datasets are freely available for many nations.

\section*{Ethics}
This paper presents a reanalysis of publicly available data. The data was used consistently with its original purpose and the analysis aggregated rather than resolved information and thus did not involve a deanonymisation risk. While the census is ultimately sourced from humans we judge it not to be personally identifiable information in the aggregated form processed here.

\section*{Data access}
The data used in this analysis is freely available and can be downloaded from www.nomisweb.co.uk/census/2011. Detailed files on the results will be made available on www.biond.org. Background maps used in the figures were generated based on data from OpenStreetMap, copyright OpenStreetMap contributors, licensed under OdBL. It is available via Open StreetMaps Overpass API.

\section*{Author contributions}
Both authors contributed equally to this paper.

\section*{Competing interests}
The authors declare that they have no competing interests.

\section*{Funding}
This work was partially supported by project Skyline under the EPSRC Encore Network+ (EP/N010019/1) and EPSRC standard grant EP/N034384/1

\section*{Acknowledgements}
The authors would like to thank Jack Booth for exploratory work during his master thesis project at the University of Bristol. TG thanks Boaz Nadler for long discussions about diffusion maps. The authors also thank the contributors to OpenStreetMap for providing an excellent open resource and Scott Hudson, of Ocean Letting and Management, for discussions of Bristol housing.


 \bibliographystyle{unsrtnat}

\bibliography{refs.bib}

\end{document}